\documentclass[prb,twocolumn,showpacs,preprintnumbers,amsmath,amssymb,floatfix,groupedaddress,superscriptaddress,aps,10pt]{revtex4-1}

\usepackage{graphicx}
\usepackage{dcolumn}
\usepackage{bm}
\usepackage{color}
\usepackage{amsmath}
\usepackage{multirow}
\usepackage[usenames,dvipsnames,svgnames,table]{xcolor}

\newcommand{\Tl}{Tl$_2$Mo$_6$Se$_6$}

\newcommand{\Na}{Na$_{2-\delta}$Mo$_6$Se$_6$}
\newcommand{\MMoSe}{$M_2$Mo$_6$Se$_6$}

\begin{document}

\title{Dimensional crossover in the quasi-one-dimensional superconductor Tl$_2$Mo$_6$Se$_6$}

\author{S. Mitra}
\affiliation{Division of Physics and Applied Physics, School of Physical and Mathematical Sciences, Nanyang Technological University, 21 Nanyang Link, Singapore 637371.}
\author{A.P. Petrovi\'c}
\email{appetrovic@ntu.edu.sg}
\affiliation{Division of Physics and Applied Physics, School of Physical and Mathematical Sciences, Nanyang Technological University, 21 Nanyang Link, Singapore 637371.}
\author{D. Salloum}
\affiliation{Sciences Chimiques, CSM UMR CNRS 6226, Universit\'e de Rennes 1, Avenue du G\'en\'eral Leclerc, 35042 Rennes Cedex, France.}
\affiliation{Faculty of Science III, Lebanese University, PO Box 826, Kobbeh-Tripoli, Lebanon.}
\author{P. Gougeon}
\affiliation{Sciences Chimiques, CSM UMR CNRS 6226, Universit\'e de Rennes 1, Avenue du G\'en\'eral Leclerc, 35042 Rennes Cedex, France.}
\author{M. Potel}
\affiliation{Sciences Chimiques, CSM UMR CNRS 6226, Universit\'e de Rennes 1, Avenue du G\'en\'eral Leclerc, 35042 Rennes Cedex, France.}
\author{Jian-Xin Zhu}
\affiliation{Theoretical Division and Center for Integrated Nanotechnologies, Los Alamos National Laboratory, Los Alamos, New Mexico 87545, USA}
\author{C. Panagopoulos}
\email{christos@ntu.edu.sg}
\affiliation{Division of Physics and Applied Physics, School of Physical and Mathematical Sciences, Nanyang Technological University, 21 Nanyang Link, Singapore 637371.}
\author{Elbert E. M. Chia}
\email{elbertchia@ntu.edu.sg}
\affiliation{Division of Physics and Applied Physics, School of Physical and Mathematical Sciences, Nanyang Technological University, 21 Nanyang Link, Singapore 637371.}
\date{\today}

\begin{abstract}
We present magnetic penetration depth and electrical transport data in single crystals of quasi-one-dimensional (q1D) Tl$_2$Mo$_6$Se$_6$, which reveal a 1D$\rightarrow$3D superconducting dimensional crossover.  The $c$-axis penetration depth shows the onset of superconducting fluctuations below $T_{1D}^{ons}=$~6.7~K, whereas signatures of superconductivity in the $ab$-plane penetration depth -- a uniquely sensitive probe of the transverse phase stiffness -- only emerge below $T_{3D}^{ons}=$~4.9~K. An anomalously low superfluid density persists down to $\sim$3~K before rising steeply, in agreement with a theoretical model for crossovers in q1D superconductors. Our data analysis suggest that a sequence of pairing and phase fluctuation regimes controls the unusually broad superconducting transition.  In particular, the electrical resistivity below $T_{3D}^{ons}$ is quantitatively consistent with the establishment of phase coherence through gradual binding of Josephson vortex strings to form 3D loops. This dimensional crossover within the superconducting state occurs despite the relatively large transverse hopping predicted from band structure. Our results have important consequences for the low-temperature normal state in Tl$_2$Mo$_6$Se$_6$ and similar q1D metals, which may retain one-dimensional behavior to lower temperatures than expected from theory.  
	
\end{abstract}

\maketitle

\section{Introduction}

Quasi-one-dimensional (q1D) materials can be regarded as arrays of parallel one-dimensional (1D) chains with a weak transverse coupling between chains.  Although long-range electronic order would be suppressed by fluctuations in a perfectly 1D material~\cite{Mermin1966,Hohenberg1967}, inter-chain hopping in q1D systems always becomes relevant at sufficiently low temperatures~\cite{Efetov1975,Scalapino1975,Gor'kov1975}.  This inter-chain hopping/coupling drives a dimensional crossover from 1D (intra-chain) to 2D or 3D (intra \emph{and} inter-chain) behavior, depending on the array anisotropy. 


Several compelling reasons exist to study such crossovers.  For example, q1D materials provide a unique experimental access to theoretically tractable models for electron-electron ($e^-$-$e^-$) correlations in higher dimensions~\cite{Gogolin1998}, but remain poorly understood~\cite{Carr2003a}. Interactions between 1D charge and/or spin stripes also influence the behavior of many complex materials, including cuprate superconductors and magnetic nickelates~\cite{Vojta2006,Anissimova2014,Comin2015}. In particular, arrays of nanoscale stripes provide an attractive method of enhancing superconducting transition temperatures via resonant confinement effects~\cite{Perali1996,Bianconi1997}. Understanding the process by which such stripes couple to form long-range ordered states is therefore of great importance.

Dimensional crossover out of a 1D state is primarily governed by the transverse electron hopping integral $t_{\perp}$~\cite{Giamarchi2003}. In the non-interacting limit, coherent single-particle inter-chain hopping occurs for temperatures lower than $t_\perp/k_B$. The presence of $e^{-}$$-$$e^{-}$ interactions renormalizes this crossover to lower temperatures, and for sufficiently strong correlations, single-particle hopping becomes irrelevant. However, a lack of single-particle hopping does not preclude the establishment of long range order, since coherent two-particle hopping eventually occurs below a characteristic temperature $T_{x2}$, controlled by a combination of the anisotropy, interaction strength and binding energy~\cite{Brazovskii1985,Firsov1985,Bourbonnais1988,Boies1995}.  This creates a ground state featuring ordered pairs, i.e. a superconductor or density wave, at a temperature $T_c$ below $T_{x2}$~\cite{Bourbonnais1991}. Two-particle crossovers are rare~\cite{Wzietek1993,Klemme1995} and largely unexplored, since most q1D materials studied to date exhibit large $t_\perp\gtrsim$~100~K and/or weak pairing instabilities, hence undergoing \textit{single}-particle dimensional crossover in their normal (metallic) states~\cite{Hussey2002,Giamarchi2004,DosSantos2007}.  In this paper, we investigate possible \textit{two}-particle mediated dimensional crossover to a superconducting ground state in the q1D material {\Tl}.

If superconductivity in a q1D material emerges directly from a normal state which is a 1D electron liquid, the superconducting transition is expected to be radically different from the narrow mean-field transitions observed in 3D materials~\cite{Efetov1975}.  The temperature at which pairing occurs within individual chains, $T_p$, is independent of the inter-chain Josephson coupling temperature $T_{x2}$ --- if $T_{x2} < T_{p}$, we anticipate a broad ``two-step'' superconducting transition~\cite{Ansermet2016}.  Fluctuating 1D superconductivity initially develops below $T_p$, but only exhibits \emph{local} intra-chain phase coherence.  As temperature is further reduced, dimensional crossover to 3D superconductivity (featuring transverse phase coherence) begins at $T_{x2}$, with a zero-resistivity Meissner state only attained at an even lower temperature $T_c < T_{x2}$. The nature of the phase fluctuations in the 1D phase, and possible topological characteristics of the phase ordering below $T_{x2}$ remain unclear. Our present work aims to clarify these issues.  

\begin{figure}[tb]
	\centering 
	\includegraphics[clip=true, width=0.99\columnwidth]{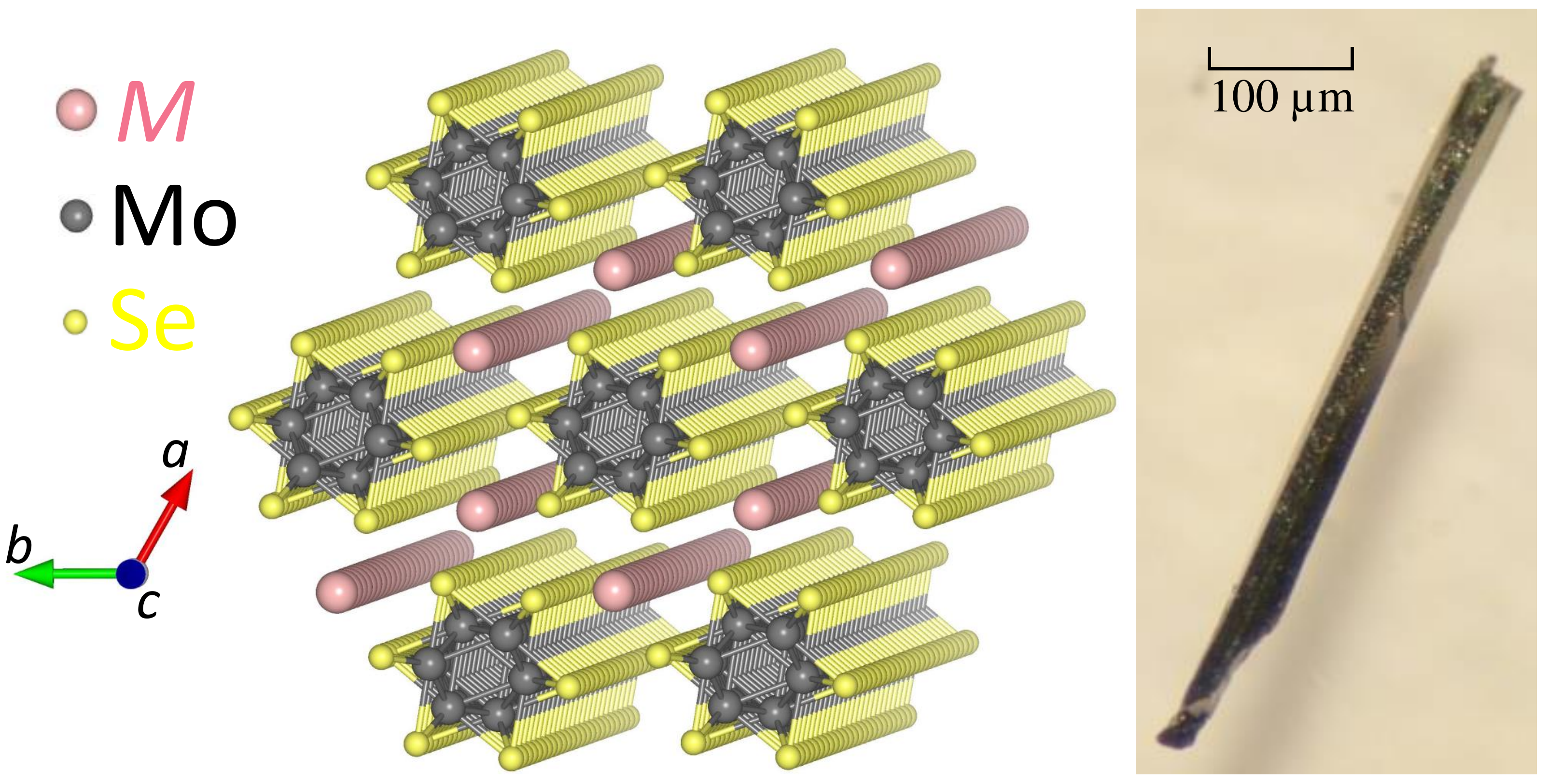}
	\caption{\label{Fig1} Crystal structure of $M_2$Mo$_6$Se$_6$, viewed at an oblique angle close to the $c$-axis. The space group is hexagonal $P6_3/m$ with the $a$ and $c$ axis lattice parameters equal to 8.94{\AA} and 4.50{\AA} respectively in Tl$_2$Mo$_6$Se$_6$.  The seven (Mo$_6$Se$_6$)$_\infty$ chains illustrated above have been cut at a length of 25 unit cells to preserve clarity.   An optical micrograph of a typical Tl$_2$Mo$_6$Se$_6$ crystal is also shown.}
\end{figure}

Recently, anomalously broad transitions have been observed in a variety of crystalline q1D superconductors~\cite{Petrovic2007,Petrovic2010,Bergk2011,He2015,Ansermet2016,Tsuchiya2017}. The {\MMoSe} family~\cite{Potel1980} ($M$ is a Group IA or IIIA cation) are of key interest, since they possess ideal 1D Fermi surfaces whose warping, and hence $t_\perp$, can be tuned by varying the $M$ ion~\cite{Petrovic2010,Petrovic2016,Liu2017}. The anisotropic crystal structure and typical geometry are shown in Fig.~\ref{Fig1}: {\MMoSe} is composed of infinite-length (Mo$_6$Se$_6$)$_\infty$ chains, oriented along the crystallographic $c$-axis.  These chains can be considered as linear condensations of the Mo$_6$Se$_8$ Chevrel-type cluster, a building block for low-dimensional short coherence length superconductors~\cite{Fischer1978,Pena2015}. The $M$ ions lie in the channels between the chains, hence facilitating inter-chain coupling.  Crucially, {\MMoSe} are isotropic in the $ab$ plane and should therefore crossover directly into a 3D state at low temperature rather than passing through an intermediate 2D regime.  

{\Tl} was the first superconducting member of {\MMoSe} to be discovered~\cite{Armici1980a}, has the highest transition temperature ($T_c\sim$~4.2~K) and does not exhibit disorder-induced localization since the Tl vacancy population is low (typically $\leq$~5\%)~\cite{Petrovic2016}.  Previous transport experiments revealed broad superconducting transitions \textcolor{red}{with an onset temperature $\sim6.7$~K,}~\cite{Petrovic2007,Petrovic2010,Bergk2011} and an unusual non-zero differential resistance plateau within the transition~\cite{Bergk2011}. Measurements in the related (yet considerably more anisotropic) {\Na} displayed evidence for 1D phase fluctuations in a similar region of the transition~\cite{Ansermet2016}.  However, the resistivity eventually falls to zero and a Meissner effect emerges, indicating the formation of a 3D phase-coherent ground state.  These observations are consistent with a two-step superconducting transition and its associated two-particle dimensional crossover.

A serious challenge to this two-step scenario is posed by DFT calculations, which yield $t_{/\!/}\sim$~12000~K, $t_\perp\sim$~230~K in {\Tl}~\cite{Petrovic2010,Ansermet2016}.  The fact that $t_\perp\approx50T_c$ implies that the low-temperature normal state \emph{should} be an anisotropic Fermi liquid, with coherent single-particle hopping between (Mo$_6$Se$_6$)$_\infty$ chains.  A sharp transition would then be expected, subject to limited broadening from 3D fluctuations~\cite{Larkin2008}. Sharp transitions are indeed observed in other q1D superconductors such as Bechgaard salts~\cite{Jerome2015} and purple bronze~\cite{Greenblatt1984} (which exhibit comparable $t_\perp\sim$~100~K) as well as chromium pnictides~\cite{Bao2015}.  It is therefore important to (a) clarify whether a true 1D$\rightarrow$3D crossover is causing the broad transitions observed in {\MMoSe} and (b) investigate the mechanism by which a 3D phase-coherent ground state is established. 

To resolve these puzzles, we track the evolution of the anisotropic phase stiffness in {\Tl} using a uniquely sensitive tunnel diode oscillator (TDO) technique.  The normalized superfluid densities $\rho^s_{ab,c}\propto\mid$$\Psi$$\mid^{2}$ are proportional to the phase stiffness and can be extracted from the measured temperature-dependent magnetic penetration depths $\lambda_{ab,c}(T)$.  Our main discovery is that inter-chain phase coherence only emerges at a temperature $T_{3D}^{ons}=$~4.9K which is lower than the onset of local intra-chain coherence at $T_{1D}^{ons}=$~6.7~K.  This supports the presence of a 1D$\rightarrow$3D dimensional crossover at $T_{x2} \equiv T_{3D}^{ons}$ in {\Tl}, in spite of the large $t_\perp$. Our observations are qualitatively consistent with a theoretical study of superfluid density within a fully microscopic effective model, which incorporates a strong anisotropy between in-plane and $c$-axis hopping integrals. Comparing penetration depth and electrical transport data suggests that the superconducting transition in {\Tl} comprises a sequence of fluctuation regimes.  Notably, the resistivity displays signatures of an exponentially-diverging correlation length below $T_{3D}^{ons}$, implying that vortex binding plays an important role in a topological 3D phase-ordering process.  

Our paper is organized as follows: in Sec. II we describe the application of our penetration depth technique to q1D superconductors, before presenting our results for the anisotropic $\lambda_{ab,c}(T)$, $\rho^s_{ab,c}(T)$ and a representative calculation for the theoretical superfluid density in a q1D material. We correlate these data with $c$-axis resistivity measurements in Sec. III, and we discuss the phase ordering below the dimensional crossover in Section V.  Our findings are summarized in Sec. V, where we briefly outline the consequences of our results for the normal state in {\Tl} and other q1D metals.

\section{Anisotropic Magnetic Penetration Depth}

TDO techniques~\cite{Prozorov2006} have previously been used to measure the superfluid density of a variety of exotic superconductors~\cite{Bonalde2000,Chia2003,Chia2005,Cho2012a}, but have rarely been employed in q1D materials~\cite{Pang2015} and never in a superconductor with a putative two-step transition.  It is therefore useful to discuss how a TDO experiment is influenced by q1D physics.  

The TDO functions as an extremely sensitive ac susceptometer, whose resonant frequency changes due to flux exclusion from a superconducting sample placed inside a detection coil.  Below the superconducting transition temperature, it can be shown that the change in the diamagnetic susceptibility $\Delta \chi(T)$ = $\chi$(\textit{T}) -- $\chi$ $(0.35~\text{K})$ is related to the change in the oscillator resonant frequency $\Delta $\textit{f}(\textit{T}) as follows:~\cite{Carrington1999}
\begin{equation} \label{ChiT}
4\pi\Delta\chi(T)=\frac{G}{R_{3D}}\Delta f(T)
\end{equation}           
which in turn gives
\begin{equation} \label{lambdaT}
\Delta \lambda (T) \ = \mathit{G}\Delta \mathit{f(T)}
\end{equation}          
where $\Delta$$\lambda$(\textit{T}) = $\lambda$(\textit{T}) -- $\lambda $(0.35~K) is the change in the magnetic penetration depth. We measure relative to $\lambda$(0.35~K) since this is the minimum temperature achievable in our apparatus.  $G$ is a calibration factor depending on the solenoid and sample geometry, which is determined by comparison with a reference Al sample~\cite{Chia2003a}, while $R_{3D}$ is the effective sample dimension estimated using a standard approach for samples with rectangular cross-section~\cite{Prozorov2000}.  Complete details of our experimental setup may be found elsewhere~\cite{Mitra2017}. 

\begin{figure}[tbp]
	\centering 
	\includegraphics[clip=true, width=0.99\columnwidth]{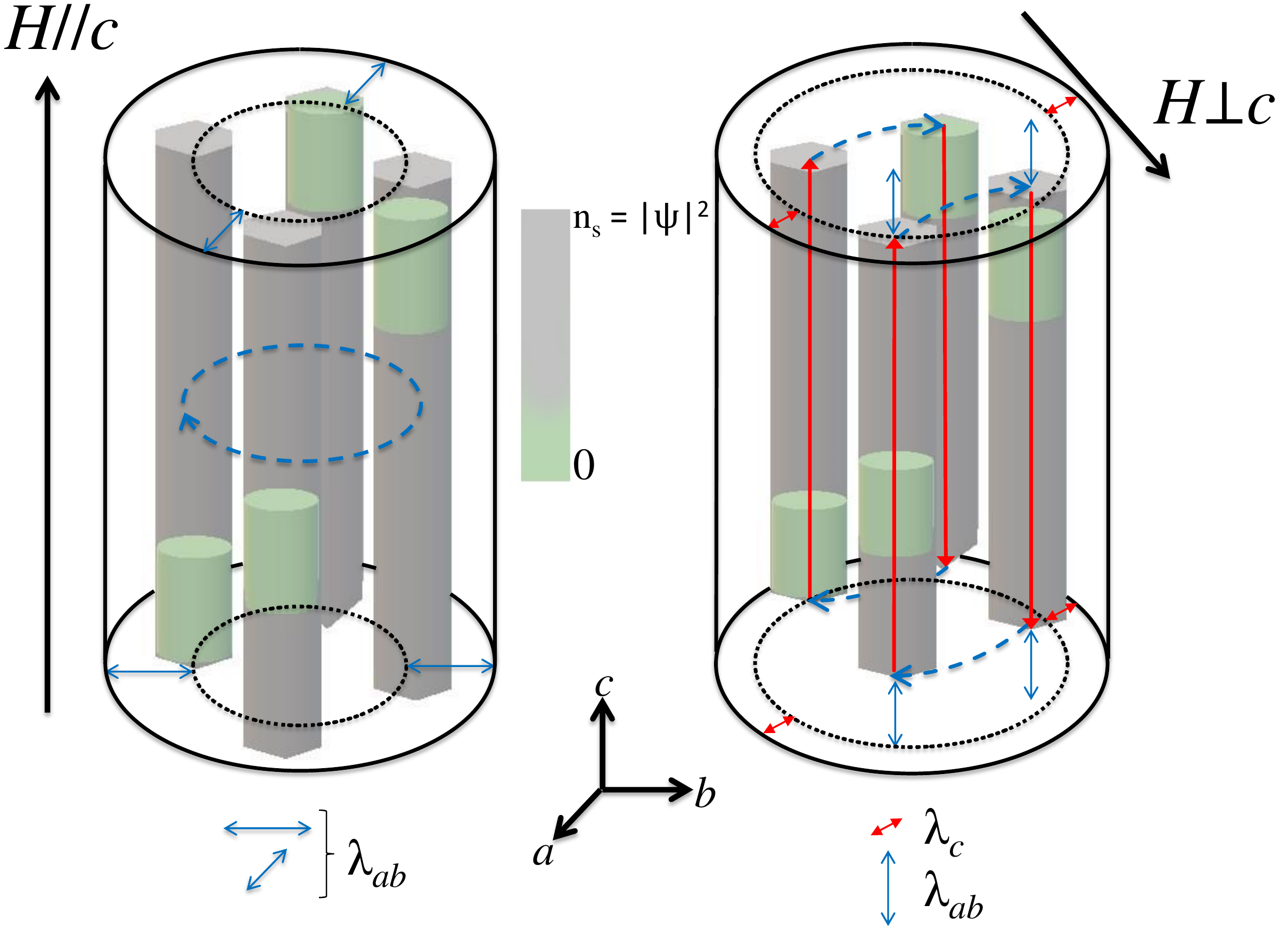}
	\caption{\label{Fig2} Schematic illustrating anisotropic flux screening in q1D superconductors.  We consider a single crystal of length $2t$ and diameter $2w$, which can be regarded as a bundle of weakly-coupled filaments.  For $H$ parallel to the filaments (left), the circulation of Meissner screening currents is entirely reliant on inter-chain Josephson coupling and hence negligible screening is anticipated for temperatures $T\gtrsim~T_{x2}$.  Flux penetrates a relatively large distance $\lambda_{ab}$ into the crystal, leading to a weakened diamagnetic susceptibility $\left|\chi\right|<$~1.  For $H$ perpendicular to the filaments (right), the screening currents flow predominantly parallel to the filaments.  In the 1D regime at $T>T_{x2}$ flux screening will be substantially weakened, due to the lack of Josephson coupling and the spontaneous formation of 1D phase slips where the amplitude of the order parameter $\left|\Psi\right|^2$ fluctuates to zero.  The diameter of the filaments and the spatial extent of a phase slip have been enlarged for clarity: phase slips occur over a lengthscale $\xi_{/\!/}\sim$~100~nm, which is 4 orders of magnitude shorter than the typical length of a {\Tl} crystal yet 2 orders of magnitude larger than the filamentary diameter.}
\end{figure} 

Figure~\ref{Fig2} illustrates the flux penetration profiles and screening currents flowing in a q1D superconductor, for magnetic fields $H$ applied parallel or perpendicular to the high symmetry $c$-axis.  For $H/\!/c$, screening is entirely due to inter-chain (Josephson) supercurrents and and the flux penetration is described by a single lengthscale $\lambda_{ab}$.  Measurements in this field configuration hence provide a highly sensitive probe of transverse phase coherence.  In contrast, for $H/\!/ab$ the screening currents comprise both intra and inter-chain components.  We therefore measure an effective penetration depth $\Delta$$\lambda_{eff}$, which contains contributions from both in-plane $\Delta\lambda_{ab}$ and out-of-plane $\Delta\lambda_{c}$.  Since $\Delta\lambda_{ab}$ can be extracted from our $H/\!/c$ dataset, we extract $\Delta\lambda_{c}$ using the relation $(\Delta\lambda_{ab}/t)+(\Delta\lambda_{c}/w)=(\Delta\lambda_{eff}/R_{3D})$ derived by Prozorov $et~al.$~\cite{Prozorov2011}, where 2$t$ and 2$w$ represent the sample length and diameter respectively.  In a q1D crystal $w \ll t$, and so $\Delta\lambda_{eff}$ will naturally be dominated by $\Delta\lambda_{c}$.  Measurements with $H/\!/ab$ are hence primarily sensitive to intra-chain coherence and phase slips.  By measuring in both field orientations and comparing the evolution of $\Delta\lambda_{ab,c}(T)$ through the transition, we can determine the dimensionality and phase stiffness of the superconducting condensate.  

\begin{figure}[tbp]
	\centering 
	\includegraphics[clip=true, width=0.99\columnwidth]{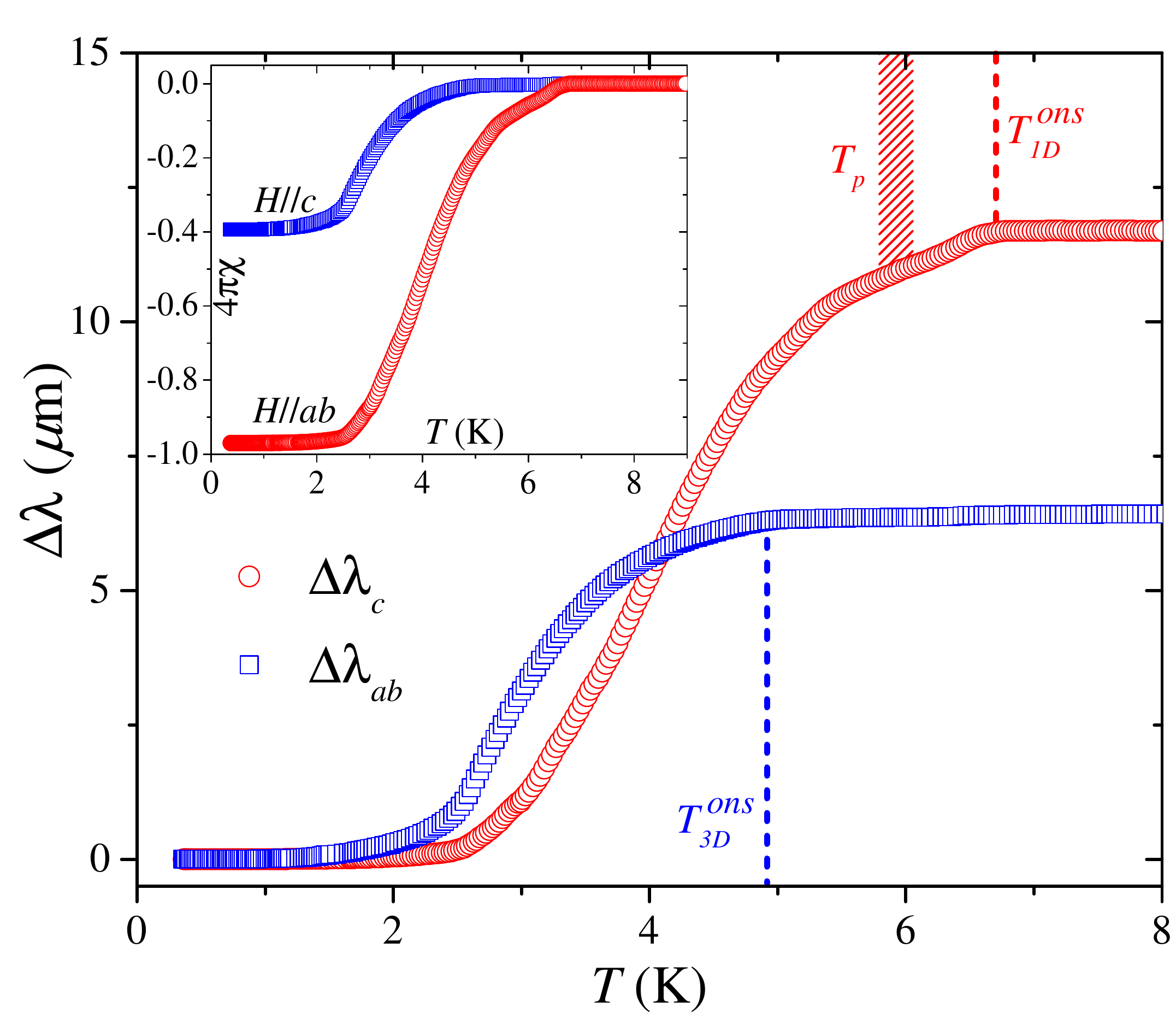}
	\caption{\label{Fig3} Anisotropic $\Delta\lambda_{ab,c}(T)$ data for Tl$_2$Mo$_6$Se$_6$, obtained by our TDO technique. Dashed lines correspond to the three key temperatures separating the various regimes within the superconducting transition: $T_{1D}^{ons}=6.7$~K (onset of the transition in $\Delta\lambda_c$), $T_p\sim$ 5.9~K (point of inflexion in $\Delta\lambda_c$) and $T_{3D}^{ons}=4.9$~K (onset of the transition in $\Delta\lambda_{ab}$). $T_{1D}^{ons}$ ($T_{3D}^{ons}$) is defined as the intersection point of the extrapolated linear regions immediately before and after $\Delta\lambda_{c}(T)$ ($\Delta\lambda_{ab}(T)$) starts to fall from its normal state value.  Close inspection reveals a very faint anomaly in $\Delta\lambda_{ab}(T)$ at $T_{1D}^{ons}$: less than 1\% of the total $\Delta\lambda_{ab}$.  This is likely due to an error in the crystal alignment with the ac excitation field, whose accuracy we estimate to be within $\pm$1\% in our apparatus. The inset shows the anisotropic magnetic susceptibilities $4\pi\chi(T)$ for $H/\!/c$ and $H/\!/ab$, obtained using Eq.~(\ref{ChiT}) with $4\pi \chi(6.75$~K) = 0.  }
\end{figure}

TDO penetration depth measurements were performed on a needle-shaped Tl$_2$Mo$_6$Se$_6$ single crystal, with length $\sim$1.5~mm and diameter $\sim$0.070~mm. The crystallographic $c$-axis corresponds to the morphological needle axis.  The crystal growth and characterization procedures have been described previously~\cite{Potel1980,Petrovic2010}. Figure~\ref{Fig3} shows $\Delta\lambda_{ab,c}(T)$ for Tl$_2$Mo$_6$Se$_6$ from 0.35~K to 9~K, extracted from the TDO resonant frequency shift $\Delta f(T)$ using Eq.~(\ref{lambdaT}), with $H_{ac}\sim50$~mOe.  It is immediately clear from the $\Delta\lambda_c(T)$ curve that the superconducting transition is unusually wide, stretching from $\sim$2.6--6.7~K, despite the small applied field. However, the observed transition width has proved robust in measurements on other {\Tl} crystals, and previous dc magnetization data in Tl$_2$Mo$_6$Se$_6$ show similarly broad transitions~\cite{Bergk2011}.  Converting $\Delta f(T)$ to diamagnetic susceptibility using Eq.~(\ref{ChiT}) yields effective superconducting volume fractions $\sim$39\% for $H/\!/c$ and $\sim$99\% for $H/\!/ab$, as shown in the inset of Fig.~\ref{Fig3}. This is in line with theoretical expectations for q1D superconductors with $\lambda_{ab}\gg\lambda_c$ as well as early magnetization experiments on Tl$_2$Mo$_6$Se$_6$~\cite{Brusetti1988a}.

The principal message from these data is that the apparent onset temperature for the superconducting transition varies with field orientation: 

\begin{itemize}
	\item $T_{1D}^{ons}=$~6.7~K for $\Delta\lambda_c(T)$ (measured with $H/\!/ab$)
	\item $T_{3D}^{ons}=$~4.9~K for $\Delta\lambda_{ab}(T)$ (measured with $H/\!/c$)
\end{itemize}

This observation is consistent with a two-step transition scenario, where (1) local intra-chain phase coherence is visible in $\lambda_{c}(T)$ at higher temperatures than the (2) Josephson-mediated inter-chain coherence that establishes 3D long-range order, and which controls $\lambda_{ab}(T)$.  The magnetic properties of the transition therefore support the presence of a fluctuating 1D regime for $T_{3D}^{ons}<T<T_{1D}^{ons}$.  In addition to providing clear evidence for dimensional crossover in {\Tl}, our data also reveal a point of inflexion in $\Delta\lambda_{c}(T)$ at $T_p\sim$ 5.9~K, whose likely origin will be discussed in section III.  
 
\begin{figure}[tbp]
	\centering 
	\includegraphics[clip=true, width=0.99\columnwidth]{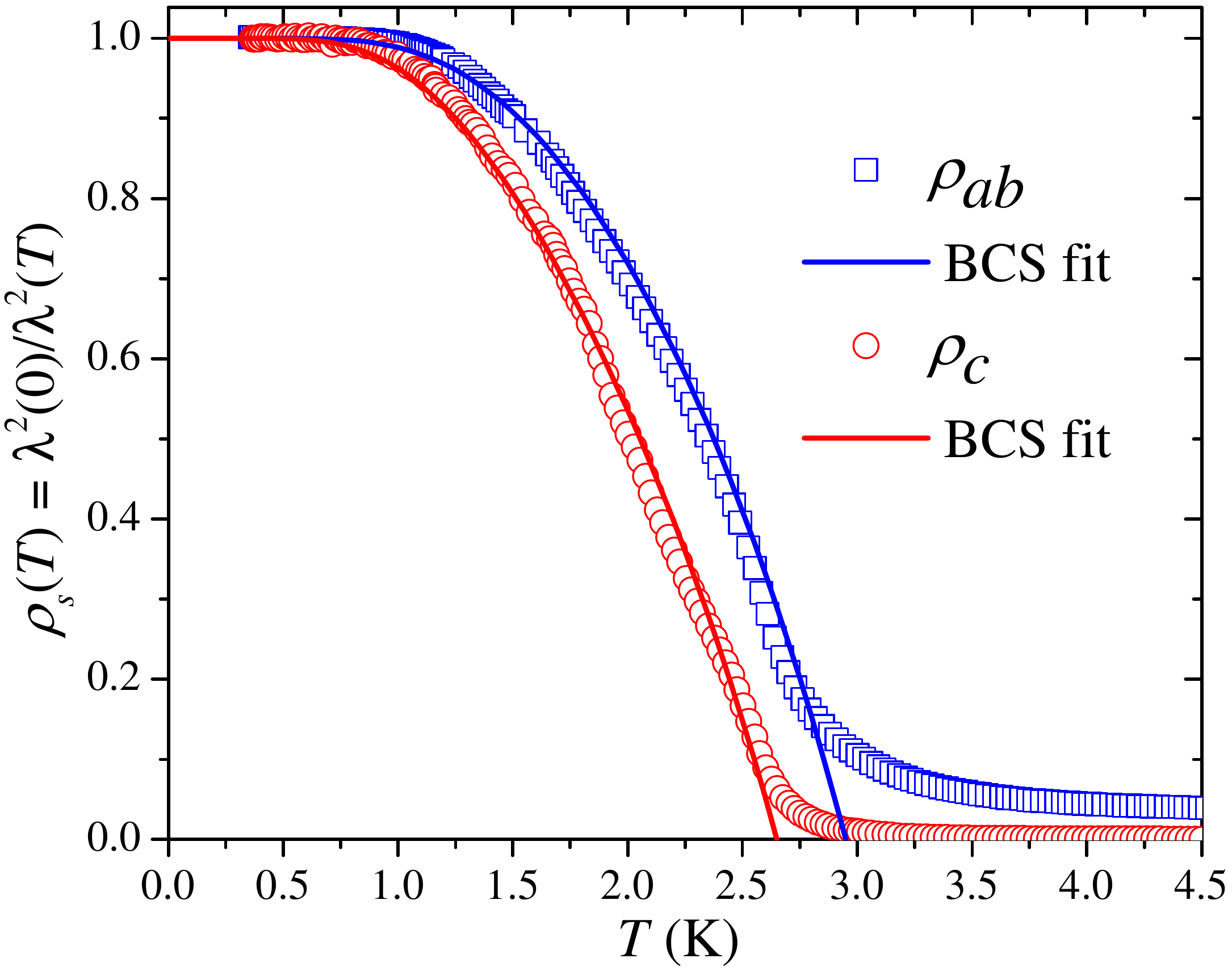}
	\caption{\label{Fig4} Normalized low temperature superfluid densities $\rho^s_{ab,c}(T)$ for {\Tl}, extracted from $\Delta\lambda_{ab,c}(T)$ in Fig.~\ref{Fig3} using $\lambda_{ab}(0)=1.5$~$\mu$m and $\lambda_{c}(0)=0.12$~$\mu$m. Solid curves are isotropic BCS $s$-wave fits to Eq.~(\ref{eqn:rhoTl}); the fit parameters are detailed in the main text.}
\end{figure}

Converting our $\Delta\lambda(T)$ data to normalized superfluid densities $\rho^{s}(T)$ = [$\lambda^{2}$(0)/$\lambda^{2}(T)$] using $\lambda_{ab}(0)=1.5$~$\mu$m and $\lambda_{c}(0)=0.12$~$\mu$m (obtained from earlier magnetic and thermodynamic data~\cite{Petrovic2010}), we plot $\rho^s_{ab,c}(T)$ in Fig.~\ref{Fig4}.  In the absence of any models specific to q1D geometry in the literature, we fit $\rho^s_{ab,c}(T)$ using a conventional 3D $s$-wave model in the clean and local limits~\cite{Tinkham}:
\begin{equation} \label{eqn:rhoTl}
\rho^{s}(T) = 1 + 2\int^{\infty}_{0} \frac{\partial f}{\partial E}
d\varepsilon,
\end{equation}     
where $f$ = [exp(E/{\it k}$_{B}T$)+1]$^{-1}$ is the Fermi function and $E = [\varepsilon^{2}$ + $\Delta (T)^{2}$]$^{1/2}$ is the Bogoliubov quasiparticle energy. We consider a Bardeen-Cooper Schrieffer (BCS) temperature dependence for the gap $\Delta(T)$ of the form~\cite{Gross1986}
\begin{equation} \label{eqn:gapGross}
\Delta(\mathit{T})=\delta
_{sc}\mathit{kT}_{c}\tanh\left\{\frac{\pi}{\delta
	_{sc}}\sqrt{a\left(\frac{\Delta C}{C}\right)
	\left(\frac{T_{c}}{T}-1\right)}\right\},
\end{equation} 
where $\delta_{sc}$ = $\Delta(0)/k_{B}T_{c}$, $a = 2/3$ and $\Delta C/C \equiv \Delta C/\gamma T_{c}$.  Our $\rho^s_{ab,c}(T)$ data below $\sim$3~K are well described by Eq.~(\ref{eqn:rhoTl}), as shown by the fitted curves in Fig.~\ref{Fig4}. The fit parameters are $\delta_{sc}=2.2$$\pm$$0.1$, $\Delta C/C=2.3$$\pm$$0.3$ and $T_{c}=(2.95$$\pm$$0.05$)~K for $\rho^s_{ab}(T)$, and $\delta_{sc}=2.0$$\pm$$0.1$, $\Delta C/C=2.2$$\pm$$0.3$ and $T_{c}=(2.65$$\pm$$0.05$)~K for $\rho^s_{c}(T)$, implying that Tl$_2$Mo$_6$Se$_6$ is a strong-coupling superconductor. The strong-coupling scenario and the obtained value of $\Delta C/C\approx2.3$ are consistent with previous specific heat measurements~\cite{Petrovic2010}.  However, the fitted mean-field transition temperatures $T_c<3$~K have no physical significance in {\Tl}, and the assumed uniform 3D model is incapable of describing the intense phase fluctuations and broad transitions inherent to q1D superconductors. $\rho^s_{ab,c}(T)$ only rise steeply below $T$$\sim$3~K, yet $\Delta\lambda_{ab,c}(T)$ display ``tails'' extending up to $T_{3D}^{ons}$ and $T_{1D}^{ons}$ for $\Delta\lambda_{ab}$ and $\Delta\lambda_{c}$ respectively (Fig.~\ref{Fig3}).  This implies that the phase stiffness is anomalously weak at higher temperatures, i.e. the ground state is prone to phase fluctuations.

\begin{figure}[tbp]
	\centering 
	\includegraphics[clip=true, width=1.0\columnwidth]{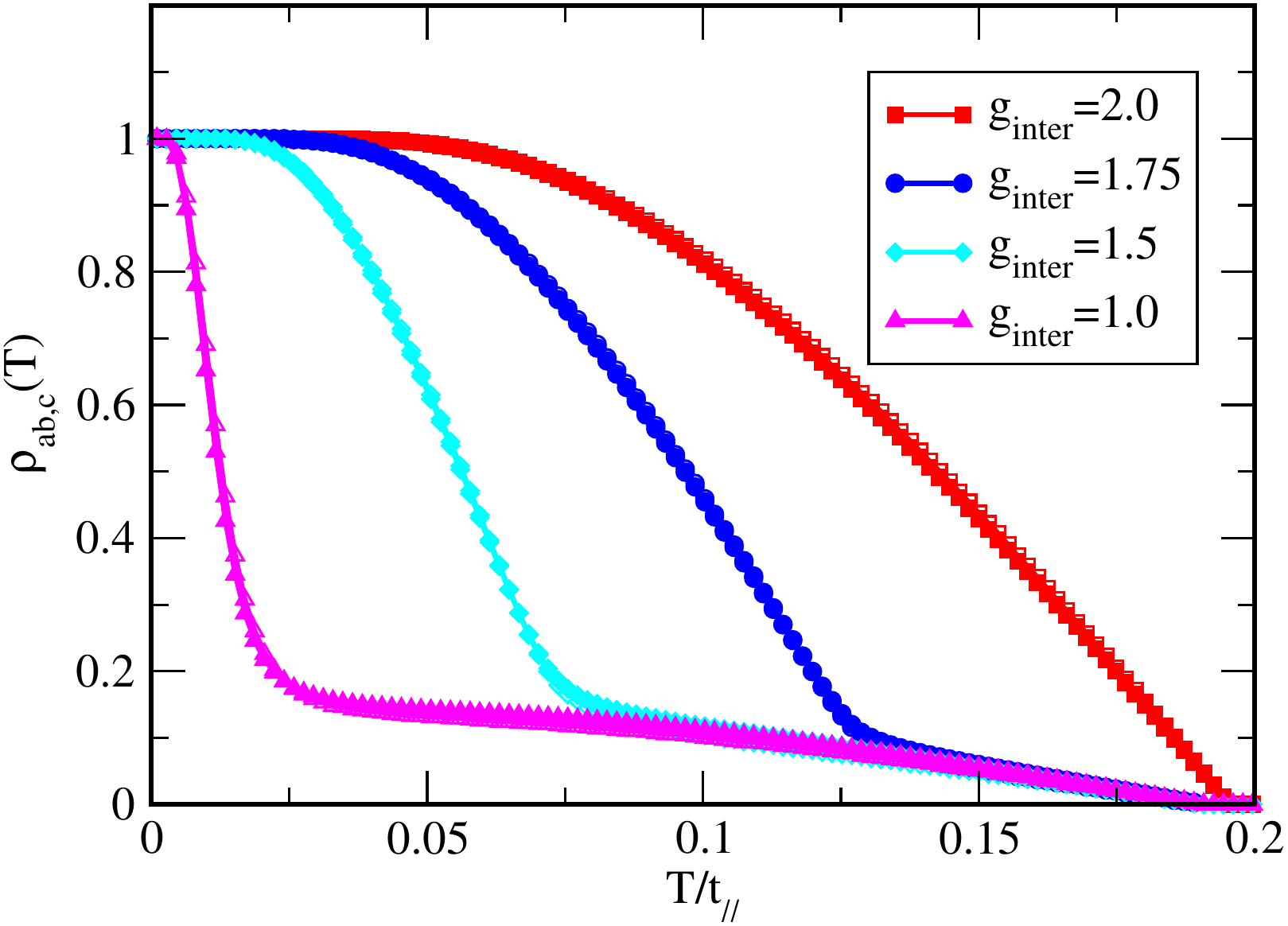}
	\caption{Calculated temperature dependence of normalized in-plane (open symbols) and $c$-axis (filled symbols) superfluid densities $\rho^s_{ab,c}(T)$ for various off-wire pairing interaction strengths $g_{inter}$, obtained within a microscopic single-band tight-binding model on a 3D cubic lattice with anisotropic nearest-neighbor hopping integrals and varying on- and off-wire pairing interactions. Our model does not consider phase fluctuations (see text for details) and cannot reproduce the anisotropic onset temperatures $T_{1D}^{ons},T_{3D}^{ons}$, so the calculated $\rho^s_{ab,c}(T)$ overlap.  In the calculations, the energy  and temperature are measured in units of $t_{/\!/}=1$.  The in-plane nearest-neighbor hopping and on-wire pairing interaction are fixed at $t_\perp=0.1$, and $g_{intra}=2$. In the $ab$-plane, the area ratio of wire to non-wire region is 4/9. The calculated ratios of $c$-axis to in-plane absolute superfluid densities near zero temperature are 186, 191, 196, and 214 for $g_{inter}$=2.0, 1.75, 1.5, and 1.0, respectively. These values lie close to the $\lambda_{ab}^{2}(0)/\lambda_{c}^{2}(0)\sim 156$ estimated from experiments~\cite{Petrovic2010}.}
	\label{FigTheory}
\end{figure}

To partially overcome the limitations imposed by a uniform 3D model, we have developed  a single-band microscopic tight-binding model for a q1D $s$-wave superconducting array. Anisotropy is introduced by varying the $ab$-plane and $c$-axis hopping integrals $t_\perp,t_{/\!/}$, as well as the pairing interactions within and between q1D chains, $g_{intra},g_{inter}$.  Within our model, the crossover temperature is controlled by the strength of a proximity effect whose energy scale is represented by the inter-wire pairing interaction $g_{inter}$ --- this corresponds to the Josephson energy in a q1D superconductor such as {\Tl}. The model Hamiltonian was diagonalized by self-consistently solving the Bogoliubov-de Gennes (BdG) equations in real space, and the superfluid density then evaluated in terms of the BdG eigenfunctions~\cite{JXZhu:2016}. Figure~\ref{FigTheory} shows the normalized in-plane and $c$-axis superfluid density. The results demonstrate that with strongly anisotropic hopping integrals and pairing interactions, the jump in $\rho^s(T)$ is suppressed to temperatures well below the onset of superconductivity.  A low plateau-like feature extends up towards the transition temperature, similar to the weak tails in our experimental $\rho^s(T)$.  These calculations are compatible with the two-step scenario exposed by our data, illustrating that the superfluid density only rises rapidly once 3D phase coherence has been established below the dimensional crossover temperature.  

It is important to note that our microscopic model does not include phase fluctuations, since these are computationally prohibitive.  For this reason, we cannot reproduce all the attributes of our experimental dataset, in particular the anisotropic onset temperatures $T_{1D}^{ons},T_{3D}^{ons}$ for $\rho^s_c,\rho^s_{ab}$.  Our low temperature data also suggest that the \emph{normalized} intra-chain phase stiffness is lower than the inter-chain stiffness, since $\rho^s_c(T)<\rho^s_{ab}(T)$ for 0.75~K~$\gtrsim T \gtrsim$~3.0~K.  This initially seems surprising, given that signatures of intra-chain coherence develop at higher temperature than inter-chain coupling ($T_{1D}^{ons}>T_{3D}^{ons}$).  We will consider the possible origins of this feature while discussing the phase ordering mechanism in Sec. IV.  

\section{Electrical Transport}

The 1D$\rightarrow$3D crossover exposed by our $\Delta\lambda_{ab,c}(T)$ data implies an important role for fluctuations below $T_{1D}^{ons}$.  We explore these by measuring the $c$-axis electrical resistivity $R_c(T)$ of the same crystal discussed in Figs.~\ref{Fig3},\ref{Fig4} using a low-frequency (19~Hz) ac four-probe method.  A set of $R_c(T)$ curves obtained for excitation currents 0.05~$\leq$~$I$~$\leq$~0.5~mA is shown in Fig.~\ref{Fig5}(a). 

For the lowest bias current $I= 0.05$~mA, the transition in {\Tl} is unusually wide, stretching from 4.4~K to 6.7~K. The existence of a broad regime of 1D fluctuations is supported by the 1D Ginzburg-Levanyuk criterion~\cite{Mishonov2003,Larkin2008}:
\begin{equation} \label{G1D}
G_{1D}=\frac{k_B}{8\sqrt{\pi}\Delta{C}\xi_{/\!/}(0)S}
\end{equation}
where we set $S$ as the cross-sectional area of a (Mo$_6$Se$_6$)$_\infty$ filament, $\Delta{C}$ is extracted from our fit to $\rho^s_c(T)$ (Fig.~\ref{Fig4}), and the Sommerfeld coefficient $\gamma=$~0.13~mJ~gat$^{-1}$~K$^{-2}$~\cite{Petrovic2010}, yielding a critical region of width $G_{1D}T_p=$~1.5~K, similar to {$\Delta T=T_{1D}^{ons}-T_{3D}^{ons}=$ 1.8~K from our data. In contrast, if we assume that the fluctuations are 3D, we obtain a critical region of width $G_{3D}T_p\sim$~0.1~mK, which is four orders of magnitude too small to explain our data~\cite{Petrovic2007}.  

Hints of multiple energy scales also emerge as bumps and points of inflexion in $R_c(T)$, unlike the sharp drops to zero seen in the resistivity of 3D superconductors.  We recall that low-dimensional or inhomogeneous superconductors can exhibit sequences of fluctuation regimes with differing physical origins or dimensionalities, even within sub-kelvin temperature ranges~\cite{Larkin2008,Glatz2011,Klemencic2017}. Superposing the key temperatures identified from $\Delta\lambda_{ab,c}(T)$ (Fig.~\ref{Fig3}) onto $R_c(T)$, several patterns emerge:

\begin{itemize}
	\item $T_{1D}^{ons}=$~6.7~K is the onset temperature for the transition seen in both $\Delta\lambda_c(T)$ and $R_c(T)$.  
	\item $T_p\sim$~5.9~K corresponds to a point of inflexion in both $\Delta\lambda_c(T)$ and $R_c(T)$.
	\item $T_{3D}^{ons}=$~4.9~K (the onset temperature for the transition seen in $\Delta\lambda_{ab}(T)$) corresponds to a kink in $R_c(T)$, i.e. a discontinuity in $dR_c/dT$.  Near this temperature, $R_c(T)$ also starts to exhibit a strong current dependence which is absent closer to $T_{1D}^{ons}$.
	\end{itemize}

\begin{figure}[tbp]
	\centering 
	\includegraphics[clip=true, width=0.99\columnwidth]{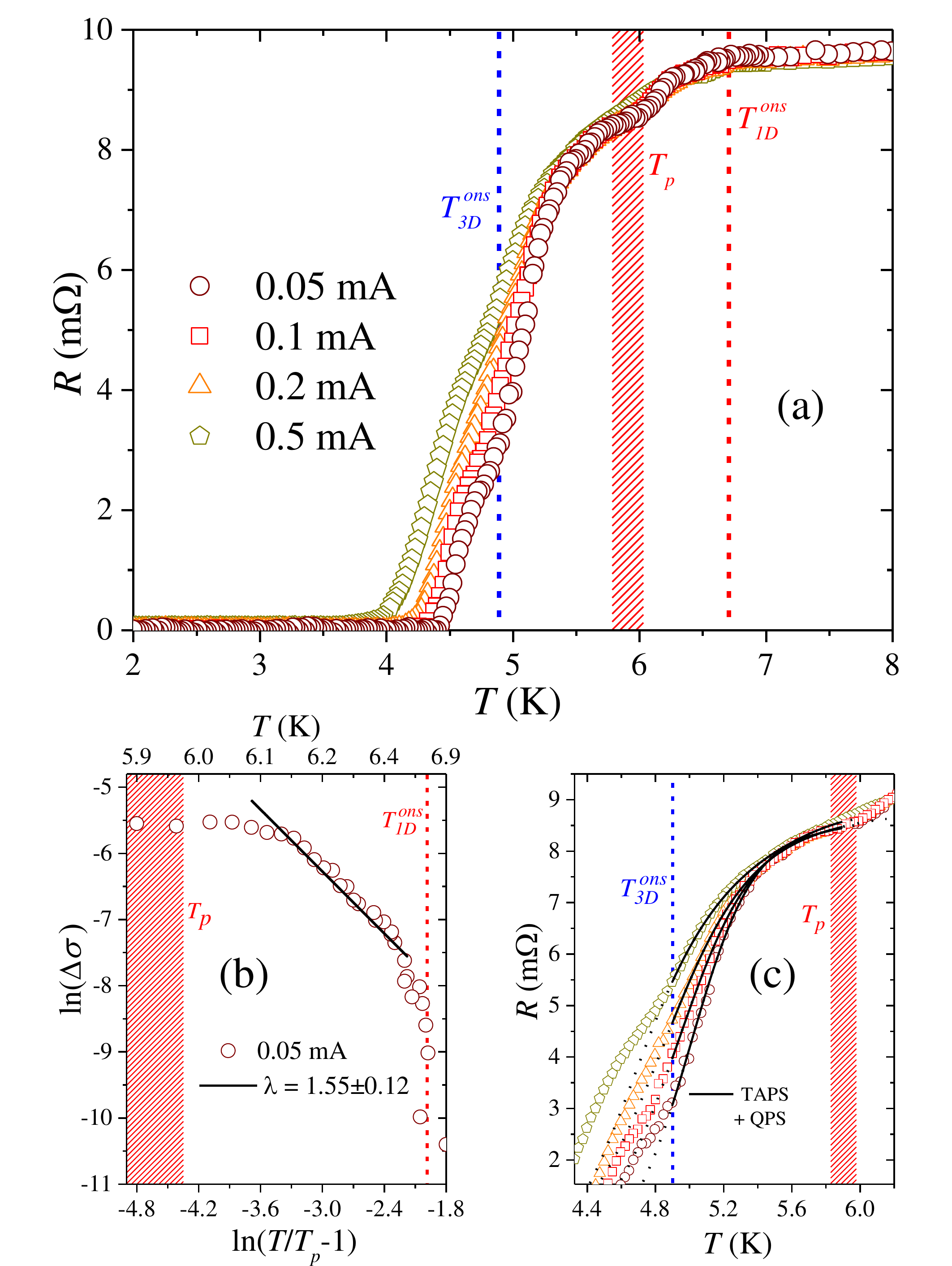}
	\caption{\label{Fig5} (a) $R_c(T)$ curves for Tl$_2$Mo$_6$Se$_6$, acquired using ac currents from 0.05-0.5~mA.  Dashed lines and shading indicate the same three temperatures identified in $\Delta\lambda_{ab,c}(T)$ (Fig.~\ref{Fig3}): $T_{1D}^{ons}$, $T_p$ and $T_{3D}^{ons}$.  (b) Aslamazov-Larkin rescaling of the paraconductivity below $T_{1D}^{ons}$.  The linear fit yields a fluctuation dimensionality D = 0.9$\pm$0.24.  Data were acquired using $I=0.05$~mA and show limited current dependence in this high temperature range.  (c) Combined thermal+quantum phase slip fits for $T<T_p$.  Dotted lines highlight the failure of the fits for $T~\leq~T_{3D}^{ons}$, heralding the onset of 3D (inter-chain) phase fluctuations.  The pairing fluctuation and 1D phase slip models are detailed in the Appendix.}
\end{figure}

The zero resistance state associated with long range order in q1D superconductors is primarily destroyed by thermal phase fluctuations, since the transverse phase stiffness is the smallest energy scale which can perturb the ground state~\cite{Efetov1975,Schulz1983a}.  However, there is no unanimously-accepted theory for the fluctuation conductivity of a q1D metal undergoing dimensional crossover into a superconducting ground state, since 1D electron liquids cannot develop order parameters with well-defined amplitude and phase.  In an attempt to circumnavigate this limitation, we analyze the different fluctuation regimes evident in $R_c(T)$ by treating {\Tl} as an array of weakly-coupled superconducting nanowires.  Upon cooling, each nanowire will first be subject to Aslamazov-Larkin (AL) pairing fluctuations~\cite{Aslamazov1968} (since the pairing energy $\Delta$ is the largest relevant energy scale), followed by 1D phase slips~\cite{Bezryadin,Altomare,Arutyunov2008} prior to dimensional crossover.  Our data reveal AL fluctuations of dimensionality $D=0.9\pm0.24$ for $T_p\lesssim{T}\lesssim{T_{1D}^{ons}}$ (Fig.~\ref{Fig5}b), and are well described by a 1D model combining thermal and quantum phase slips for $T_{3D}^{ons}\lesssim{T}\lesssim{T_p}$ (see Appendix for details and fit parameters).  We estimate a total crystal cross-section $X=$~2.5$\times$10$^{-9}$m$^2$ from the AL scaling, in good agreement with the measured cross-sectional area $\sim$3.8$\times$10$^{-9}$m$^2$ considering that current flow through q1D metals is rarely homogeneous due to microscopic cracks and disorder~\cite{Ansermet2016}. Furthermore, the pairing temperature obtained independently from the phase slip fitting lies consistently in the 5.9--6.1~K range, supporting our identification of $T_p\sim$~5.9~K as the pairing temperature in {\Tl}.  Our analysis leads us to a striking conclusion: despite its homogeneous monocrystalline nature, above $T_{3D}^{ons}$, {\Tl} experiences strong 1D fluctuations similar to a bundle of weakly-coupled superconducting nanowires~\cite{Lortz2009,Wang2010,He2013,Wong2017}.

As the temperature falls below $T_{3D}^{ons}\sim$~4.9~K, the phase slip fits fail due to a pronounced ``hump'' in $R_c(T)$.  Similar humps have previously been observed in a variety of q1D and 2D superconductors as well as Josephson junction arrays; they are usually interpreted as a combination of inhomogeneity and pair-breaking effects in materials where phase coherence is established via a vortex-binding transition\cite{Ansermet2016,Resnick1981,Humps1,Humps2}. As we will demonstrate in section IV, our data indeed provide further evidence for the emergence of vortices in {\Tl} for $T~\lesssim~T_{3D}^{ons}$. We also point out that $R_c(T)$ develops a strong current dependence below $T\approx$~5.4~K, which is accentuated by the hump below $T_{3D}^{ons}$. The enhanced effects of an applied current correspond to the transition from robust pair fluctuations to fragile phase excitations of increasing topological complexity.  

\section{Phase Ordering at the Dimensional Crossover}
Having provided evidence for 1D superconducting fluctuations below $T_{1D}^{ons}$, we finally examine the establishment of 3D superconductivity below $T_{3D}^{ons}$.  We consider {\Tl} as an array of Josephson-coupled superconducting chains, which can be modelled within a coarse-grained lattice approximation as follows~\cite{Benfatto2007}: 
\begin{equation} \label{3DXYHam}
H=\sum_{\left\langle ij\right\rangle}J_{ij}\mathrm{cos}(\theta_i-\theta_j)
\end{equation}
where $\left\langle ij\right\rangle$ denotes a sum over neighbouring lattice sites $i,j$ and $0\!<\!\theta_i\!<\!2\pi$ is the phase of the superconducting wavefunction at site $i$ which satisfies $XY$ symmetry.  $J_{ij}$ is an anisotropic energy scale describing the phase stiffness: for $i,j$ within the same chain, $J_{ij}\!\equiv\!J_{/\!/}\!\propto\!\rho^s_c$, whereas $J_{ij}\!\equiv\!J_{\perp}\!\propto\!\rho^s_{ab}$ for $i,j$ on neighbouring chains.  In the limit $J_{\perp}\!\gg\!J_{/\!/}$, we recover the well-known 3D~$XY$ model which describes phase transitions in many q2D materials~\cite{Chattopadhyay1994}.  In contrast, $J_{\perp}\!\ll\!J_{/\!/}$ for q1D superconductors and so the anisotropy is reversed.  In this case, the model still exhibits two characteristic types of excitations: long wavelength phase fluctuations and vortex loops~\cite{Shenoy1995}.  The long wavelength fluctuations only become relevant at low temperatures well below the phase ordering, and we will not discuss them further here.  As the temperature rises, small vortex loops thermally fluctuate into existence: their average diameters increase with temperature.  At the phase transition temperature $T_c$, the largest loops ``blow out'' to form free vortex strings and transverse phase coherence is lost.  

\begin{figure}[tbp]
	\centering 
	\includegraphics[clip=true, width=0.99\columnwidth]{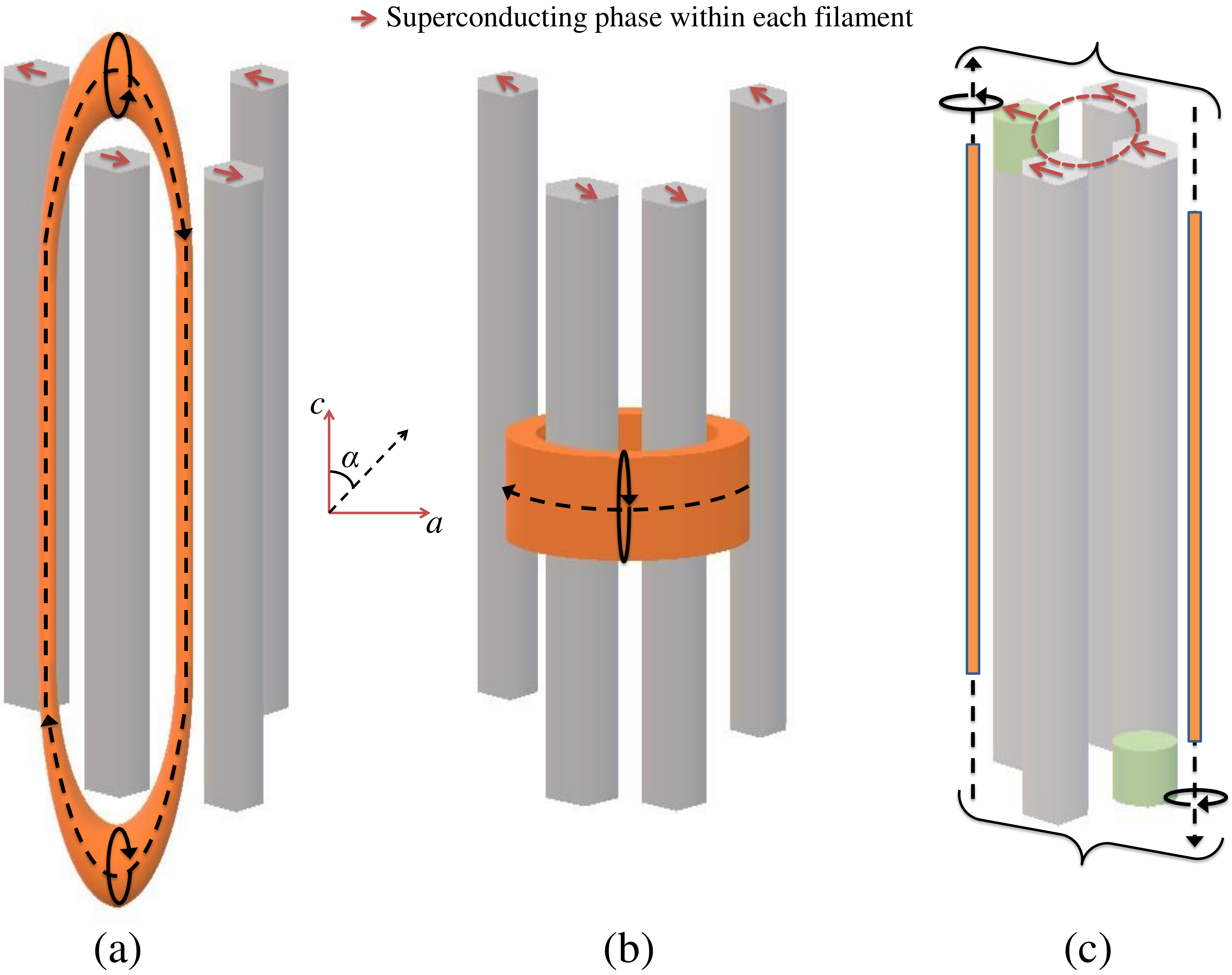}
	\caption{\label{Fig6} Vortex loop formation in a q1D superconductor.  Dashed black lines indicate the flux orientation and solid black lines correspond to the screening current flow.  (a) High fugacity elliptical loops oriented parallel to the $c$-axis.  (b) Low fugacity circular loops lying in the $ab$-plane ($\alpha=\pi/2$).  (c) 1D phase slips (green zones) act as sources/sinks for vortex segments at $T>T_c$, thus preventing any divergence in the core energy for long $c$-axis vortex strings.  As temperature is reduced, phase slip incidence falls and the vortex strings are bound into loops.}
\end{figure}

Vortex loops in a q1D superconductor are ellipses, whose axes are oriented at an angle $\alpha$ to the $c$-axis.  The two limiting cases $\alpha=0,\pi/2$ are depicted in Fig.~\ref{Fig6}(a,b): for $\alpha=0$ the major axis of the ellipse is parallel to $c$, whereas the loop forms a circle in the $ab$-plane for $\alpha=\pi/2$.  The vortex fugacity, i.e. the thermodynamic propensity towards vortex formation, will be maximal for $\alpha=0$, since the core energy is smaller by a factor $\mathcal{O}(\lambda_{ab}^2/2\lambda_c^2)\approx78$ in this orientation~\cite{Shenoy1995,Petrovic2010}.  The eccentricity of the ellipse (i.e. the ratio between the major and minor axes) can be approximated by the anisotropy parameter $\lambda_{ab}/\lambda_c\equiv\xi_{/\!/}/\xi_\perp\approx13$.  This large anisotropy implies that the majority of each vortex loop or string will lie parallel to the $c$-axis.  Above $T_c$, the length (and hence energy) of the vortex strings is prevented from diverging by 1D phase slips, as shown in Fig.~\ref{Fig6}(c): each slip changes the phase by 2$\pi$ and can hence act as a source or sink for a string.  The 3D phase fluctuations in the critical region $T_c<T<T_{3D}^{ons}$ are therefore controlled by the interactions and binding between vortex strings which are predominantly oriented along the $c$-axis.  

The interaction potential between two antiparallel vortex strings diverges logarithmically with their separation, just like the interaction between vortices and antivortices in the 2D $XY$ model leading to the Berezinskii-Kosterlitz-Thouless transition. This logarithmic dependence creates an exponential divergence in the correlation length, which results in an exponential dependence of the electrical resistivity above the phase ordering temperature.  In a q1D superconductor, exponential behavior will only be visible within a narrow temperature range above $T_c$, with a lower cut-off due to finite size effects limiting the divergence of the phase correlation length $\xi_\perp$~\cite{Holzer2001}.  Two distinct finite size effects may play a role here: $\xi_\perp$ exceeding either the crystal dimensions, or the maximum possible lengthscale for vortex-vortex interactions 2$\lambda_{ab}$.  The finite interaction length 2$\lambda_{ab}$ is likely to influence vortex dynamics in macroscopic {\Tl} crystals.  

\begin{figure}[htbp]
	\centering 
	\includegraphics[clip=true, width=0.9\columnwidth]{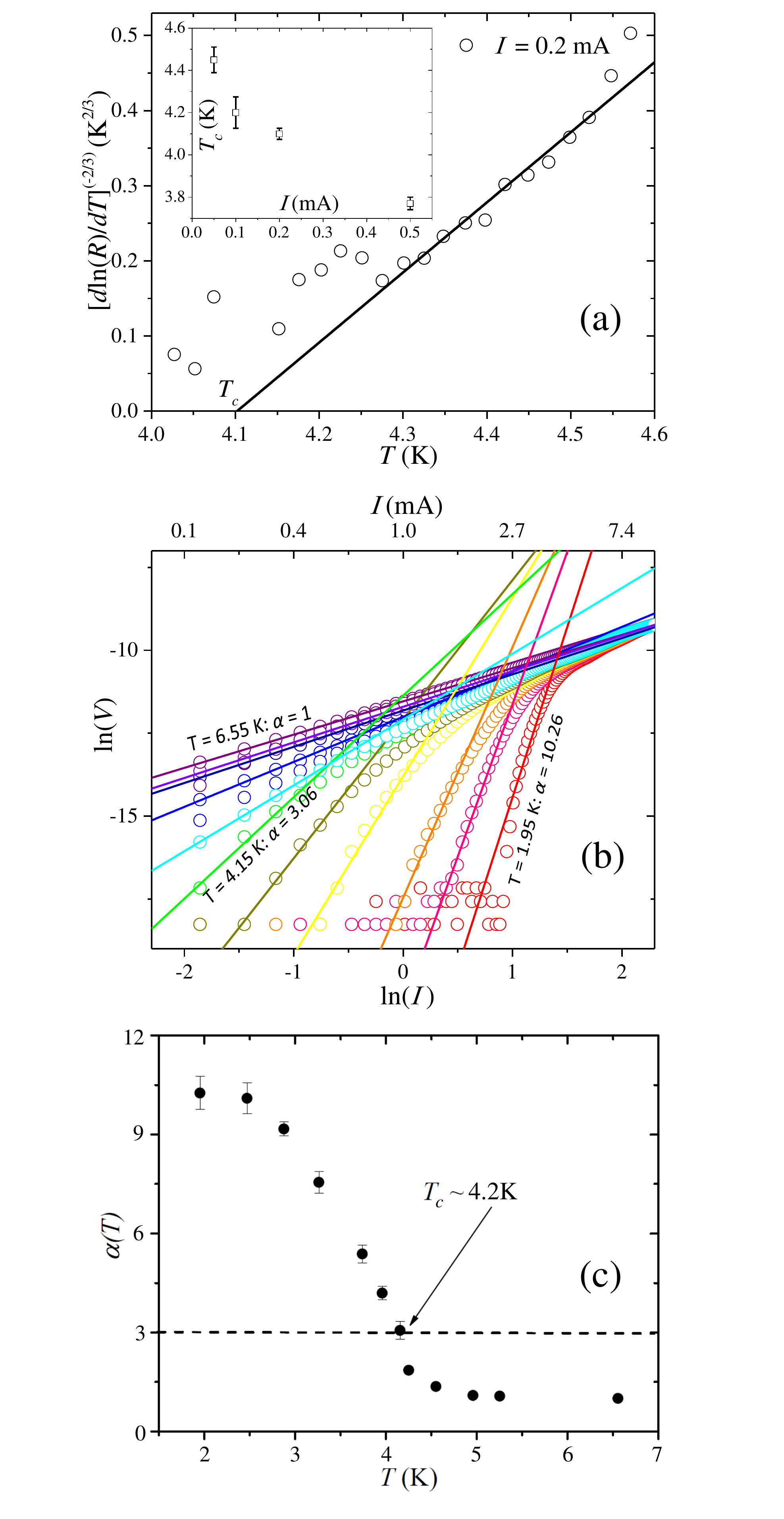}
	\caption{\label{Fig7} (a) Exponential scaling in $R_c(T)$ below $T_{3D}^{ons}$ extends over a range $\sim$~0.25~K (data obtained using $I=0.2$~mA). Extrapolating the linear fit (red solid line) to zero yields a phase-ordering temperature $T_c\approx4.1$~K. The inset shows $T_c(I)$ obtained in a similar manner for $I$ = 0.05-0.5~mA.  (b) $V(I)$ curves for Tl$_2$Mo$_6$Se$_6$ from $T=$ 1.95~K to 6.55~K, plotted on logarithmic axes.  Solid lines are linear fits corresponding to power-law behavior of the form $V$$\propto$$I^{\alpha(T)}$. The saturation of the curves at low voltage originates from a 10~nV noise threshold in our apparatus.  (c) Temperature evolution of the power-law exponent $\alpha(T)$. The estimated phase-ordering temperature $T_c \equiv T(\alpha=3)\approx4.2$~K.}
\end{figure}

In Fig.~\ref{Fig7}(a), we demonstrate the presence of exponential behaviour in $R_c(T)$ over a temperature range $\sim$~0.25~K below $T_{3D}^{ons}$, using the $R(T)\sim\exp{(b/(T-T_c))^{1/2}}$ scaling (where $b$ is a material constant) originally predicted for 2D superconductors~\cite{Halperin1979}. Similar signatures of an exponentially diverging transverse correlation length were previously observed in 4{\AA} carbon nanotube composites~\cite{Wang2010} as well as {\Na}~\cite{Ansermet2016}, and additionally seen in Monte-Carlo simulations~\cite{Sun2012}. We therefore believe this behavior to be a reproducible and generic signature of vortex-mediated phase ordering: as $\xi_\perp$ diverges, vortex/anti-vortex strings are bound into increasingly large loops, thus establishing global phase coherence.  Extrapolating this exponential regime to $R_c=0$ allows us to estimate the temperature $T_c$ at which phase order is established in the $ab$ plane, and a 3D superconducting ground state is achieved~\cite{Schneider2009}.  The inset to Fig.~\ref{Fig7}(a) shows that $T_c$ falls as the measurement current rises, suggesting that elevated currents break the vortex loops and suppress phase ordering.  

We anticipate that sufficiently large currents may break vortex loops even at temperatures below the dimensional crossover, leading to a characteristic power-law ``rounding'' in $V(I)$ curves for {\Tl}~\cite{Kadin1983}.  Figure~\ref{Fig7}(b) shows $V(I)$ data for the same crystal, plotted on a log-log scale to highlight the $V \sim I^\alpha$ power-law behavior  We plot the temperature-dependent exponent $\alpha(T)$ in Fig.~\ref{Fig7}(c): $\alpha\approx$~1 at high temperature, corresponding to the expected Ohmic behaviour for a metal, but begins to rise smoothly below $T_{3D}^{ons}\sim$~4.9~K.  In 2D materials $\alpha(T)$ jumps sharply at $T_c$, defined as $T(\alpha=3)$: this is the well-known Nelson-Kosterlitz jump in the superfluid density~\cite{Nelson1977}.  However, in materials featuring 3D vortex loops rather than 2D vortex pairs (e.g. q1D and q2D superconductors), the discontinuity at $\alpha=3$ is expected to be smeared into a gradual increase~\cite{Benfatto2007}, although $T(\alpha=3)$ should still provide a useful estimate for $T_c$.  Furthermore, outside the 2D limit $\alpha(T)$ is no longer expected to scale with the superfluid density, but instead characterizes the short-lengthscale vortex binding strength~\cite{Chattopadhyay1994}.  Our data are entirely compatible with this scenario: $\alpha(T)$ climbs smoothly through $T(\alpha=3)=4.2$~K, within the range 3.8~K~$<T_c<4.4$~K indicated by the exponential scaling in $R_c(T)$. Our observations of power-law $V(I)$ scaling and an exponential resistivity regime constitute two independent signatures of the role of vortices in establishing phase coherence at the dimensional crossover.  As the temperature is reduced further, $\alpha(T)$ only begins to saturate below $T\approx$~3~K, corresponding to the approximate temperature where $\rho^s_{ab,c}(T)$ begin to rise (Fig.~\ref{Fig4}).  This confirms that phase coherence remains fragile and topological phase excitations persist down to $\sim$~3~K, well below the dimensional crossover.    

An additional factor likely to contribute to the anomalously low superfluid density for $T\gtrsim$~3~K is the ratio of $\lambda_{ab}$ to the crystal dimensions. Vortex binding transitions in superconductors may only be observed when $2\lambda\gtrsim$ the sample width: if this condition is not fulfilled, vortices are unable to interact at large length scales and so a finite population of unbound vortices/antivortices may persist even below the nominal phase-ordering temperature~\cite{Holzer2001}.  Weak pinning of the vortices (e.g. by inhomogeneity or disorder) still allows a zero resistance state to be established, albeit with a low critical current.  In {\Tl}, $\lambda_{ab}(T)$ is of a similar size to the crystal \emph{width} (70~$\mu$m) close to $T_c$ and so vortices can interact and effectively bind in the $ab$ plane throughout the majority of the crystal cross-section. However, vortex interactions in the $ac$ or $bc$ planes are distance-limited since $\lambda_{ab}$ is much smaller than the crystal \emph{length} (1.5~mm).  This could explain the lower apparent transition temperature for $\rho^s_c(T)$ compared to $\rho^s_{ab}(T)$ from our BCS fitting (Fig.~\ref{Fig4}).  Although the absolute $\rho^s_c\propto1/\lambda_c^2~\gg~\rho^s_{ab}\propto1/\lambda_{ab}^2$, the fact that vortices cannot interact and bind along the $c$-axis over distances approaching the crystal length may reduce the $c$-axis phase stiffness and hence the normalised $\rho^s_c$.  Our data therefore hint at the existence of unpaired, weakly-pinned vortex strings at zero magnetic field following dimensional crossover in macroscopic q1D superconductors.  

\section{Conclusions}

Magnetic penetration depth measurements in q1D {\Tl} are consistent with a two-step superconducting transition, in which local phase coherence within individual 1D superconducting filaments is established at a higher temperature than global (inter-chain) coherence, i.e. $T_{1D}^{ons}>T_{3D}^{ons}$.  A 1D$\rightarrow$3D dimensional crossover therefore occurs within the superconducting state. The measured superfluid density remains small at temperatures well below the onset of superconductivity before rising steeply at lower temperature, consistent with our calculations using a q1D microscopic model.  Electrical transport measurements provide further support for {1D superconductivity above $T_{3D}^{ons}$, and suggest that homogeneous q1D crystals exhibit a similar sequence of fluctuation regimes to those expected in nanowire arrays.  We identify $T_{3D}^{ons}$ as the two-particle mediated dimensional crossover temperature $T_{x2}$, i.e. below $T_{3D}^{ons}$, power-law $V(I)$ curves and an exponential resistivity regime indicate that global phase coherence is established via a topological process in which $c$-axis vortex strings are bound to form 3D loops.  

The broad two-step transitions which we observe originate from the extreme uniaxial anisotropy of the crystal structure in {\Tl} rather than any extrinsic mechanism.  In particular, inhomogeneity or random disorder (which are common causes of anomalous broadening in superconducting transitions) can be excluded.  Macroscopic inhomogeneity cannot be reconciled with our single-gap BCS fits to $\rho^s_{ab,c}(T)$, the smooth evolution in $\Delta\lambda_{ab,c}(T)$ or the weak current dependence of $R(T)$ close to $T_{ons}$.  Moreover, random effects are inconsistent with the robust observation of similar two-step transitions in multiple {\Tl} crystals (and {\Na}) spanning a decade~\cite{Petrovic2007,Petrovic2010,Bergk2011,Ansermet2016}.

Our discovery of a two-particle superconducting dimensional crossover within the superconducting state implies that coherent single-particle inter-chain hopping is suppressed in the normal state.  Several causes for such a suppression may be envisaged.  Firstly, the true inter-chain hopping integral $t_\perp$ could be far smaller than that predicted by DFT calculations, and/or the $e^--e^-$ interactions much stronger. However, the dimensional crossover temperatures in {\Tl} and {\Na} lie close to the DFT-calculated values of $t_\perp^2/t_{/\!/}$~\cite{Ansermet2016}, which provides a good estimate for the Josephson coupling temperature in an anisotropic superconductor~\cite{Little1971,Jerome1980} -- this suggests that the calculated band structures are accurate. Alternatively, a gapped phase may be developing in the normal state.  Although we can eliminate any static Peierls ordering~\cite{Liu2017} since the low temperature crystal structure of {\MMoSe} is identical to that at room temperature~\cite{Petrovic2016}, it remains unclear whether a dynamic density wave could quench single-particle inter-chain hopping while maintaining the observed $T$-linear metallic resistivity~\cite{Petrovic2010}. Another possibility which merits consideration is the opening of a high temperature spin gap, which would gap the single-particle excitation spectrum while preserving metallic charge transport. We hope that our results will stimulate a search for hidden order within the normal state of {\MMoSe}, which has long been hypothesized~\cite{Brusetti1994a} yet never observed.

\textcolor{red}{Very recently, {\Tl} is firmly back in the spotlight thanks to an exciting prediction of topological superconductivity~\cite{q1DTSC} as well as the fascinating possibility of simulating the topological phase-ordering within a lattice of 1D fermion tubes. Such cold atom systems could  be used to test the breakdown in 1D Tomonaga-Luttinger theory and concomitant dimensional crossover due to finite transverse coupling.~\cite{Giamarchi2017} In this light, our rigorous experimental contributions to this largely unexplored field appear particularly timely and important.} 

Finally, our work indicates that in addition to showcasing a highly-unusual dimensional crossover from 1D to 3D superconductivity, q1D superconductors such as {\Tl} provide a unique environment to study topological phase ordering in systems obeying $XY$ symmetry. Here, comparison with theory and experiment in analogous cold atom models may prove especially fruitful~\cite{Iskin2009}.  The negligible superfluid density which we observe from $\sim$3K--4.9K furthermore suggests that vortex loops and strings persist over a wide temperature range within the superconducting state.  Although spatially imaging these topological defects remains a considerable challenge in q1D materials, we suggest that the combination of weak phase stiffness and uniaxial symmetry could render {\Tl} an ideal candidate for laboratory simulations of 1D cosmic string formation/propagation via the Kibble-Zurek mechanism~\cite{Williams1999}.     

\begin{acknowledgments}
We are grateful to T. Giamarchi, C. Berthod,  R. Lortz, N. Proukakis and D. Chernyshov for enlightening discussions, and to D. Ansermet for critically reading our manuscript. We acknowledge funding from the Singapore Ministry of Education (MOE) Tier 1 (RG13/12) and Tier 2 (MOE2015-T2-2-065) grants, and the Singapore National Research Foundation (NRF) Investigatorship (NRF-NRFI2015-04). This work was supported in part by the Center for Integrated Nanotechnologies, a U.S. DOE Office of Basic Energy Sciences user facility.
\end{acknowledgments}

\section{Appendix}

\subsection{Aslamazov-Larkin pairing fluctuations}
Thermally-induced pairing fluctuations in a superconductor create a paraconductivity described by the Aslamazov-Larkin (AL) model \cite{Aslamazov1968,Sharma1995}:
\begin{equation} \label{ALfluctuation}
\Delta\sigma=A\left(\frac{T}{T_c}-1\right)^{\lambda}
\end{equation}
where $\Delta\sigma = \sigma(T)-\sigma_{N}(T)$ is the excess conductivity relative to the normal state, $A$ is a constant, $T_c$ is the mean-field transition temperature and $\lambda$ describes the dimensionality $D$ of the superconducting phase, with $\lambda=2-D/2=1.5$ for 1D AL fluctuations. $\sigma_{N}(T)$ is obtained from a linear fit to $\sigma(T)$ from 7-10~K and $A=e^2\xi_0/\hbar~X$ can be evaluated from the crystal cross-sectional area $X$.  

The concept of a mean-field $T_c$ has no meaning in a two-step superconducting transition where pairing is established before phase coherence.  Determining the appropriate temperature to use as $T_c$ in an AL fit to our data is therefore not obvious: we must identify the highest temperature at which electrons are paired and locally phase coherent within individual (Mo$_6$Se$_6$)$_\infty$ chains.  Below this temperature, the transition is broadened by intra-chain phase slips and (eventually) inter-chain phase fluctuations.  The point of inflexion $T_p=$~5.9~K is an appealing candidate for this crossover between pairing and phase fluctuations: this choice is eventually justified by our observation of AL scaling above $T_p$ and 1D phase slips at lower temperature. 

We first determine the presence and dimensionality of AL fluctuations by setting $T_c~\equiv~T_p$ in Eqn.~\ref{ALfluctuation}, then plotting $\mathrm{ln}(\Delta\sigma)$ vs. $\mathrm{ln}(T/T_p-1)$ (Fig.~\ref{Fig5}(b)).  $\mathrm{ln}(\Delta\sigma)$ falls rapidly as $\mathrm{ln}(T/T_p-1)\rightarrow0$, which we attribute to short wavelength pair fluctuations controlling the paraconductivity in the high temperature limit~\cite{Reggiani1991}.  A broad linear regime emerges below 6.6~K with a slope $\lambda$ = 1.55$\pm$0.12, corresponding to dimensionality $D=0.9\pm0.24$.  Approaching $T_p=$~5.9~K, $\mathrm{ln}(\Delta\sigma)$ saturates, implying a change in the nature of the fluctuations, i.e. the impending crossover to a phase-slip dominated regime.  Pairing fluctuations should also influence {\Na}, but in this material the localization-induced divergent normal state resistance~\cite{Ansermet2016} masks the emergence of any AL component.

\subsection{Phase slips in q1D superconductors}
Phase slips are topological excitations unique to 1D superconductors, in which the amplitude of the order parameter $\left|\Psi\right|^2$ fluctuates to zero over a lengthscale $\xi_{/\!/}(T)$ with a concomitant ``jump'' in the phase by 2$\pi$~\cite{Bezryadin,Altomare,Arutyunov2008}. They are caused by thermal activation (TAPS) or quantum fluctuations (QPS), with thermal effects vanishing as $T\rightarrow0$. 

Recently a generalized thermally-activated phase slip (TAPS) theory has been successfully used to model $R_c(T)$ data in macroscopic crystals of the q1D superconductor Na$_2$Mo$_6$Se$_6$~\cite{Ansermet2016}. In their model, the authors considered the crystal as a $m \times n$ array of identical parallel 1D filaments/nanowires, each of length $L$. This leads to a geometric renormalization of $L$ to $L_{eff}=Lm/n$, where $Lm$ is the experimental voltage contact separation on a crystal and $n$ is the typical number of 1D filaments within the crystal cross-section.  Quantum phase slips (QPS) in q1D superconductors may be treated in a similar manner.  Although the exact nature of the crossover from TAPS to QPS is still under debate~\cite{Tian2005,Arutyunov2008}, models combining parallel TAPS and QPS contributions have been shown to reproduce superconducting transitions in q1D materials over broad ranges of temperature and disorder~\cite{Petrovic2016,Zhang2017}. We briefly outline such a model and its application to {\Tl} below.
  
\subsubsection{Langer-Ambegaokar-McCumber-Halperin (LAMH) model for thermally activated phase slips}

A thermally activated phase slip must overcome an energy barrier $\Delta F$, proportional to $\xi(T)=\xi(0)(1-T/T^*)^{-1/2}$ and the length of the nanowire $L$. In a nanowire made from a conventional 3D superconductor, $T^*$ would correspond to the bulk mean-field transition temperature, whereas in the q1D superconductors which we are considering $T^*$ is the pairing temperature.  The frequency of random excursions in the superconducting order parameter is given by a prefactor $\Omega(T)$ which sets the time scale of the fluctuations. The LAMH contribution to the total resistance can be expressed as 
follows~\cite{Langer1967,McCumber1970}:  
\begin{equation} \label{LAMH1}
R_{LAMH}(T)=\frac{\pi \hbar^{2}\Omega}{2e^{2}k_{B}T}\exp\left( -\frac{\Delta F}{k_{B}T}\right),
\end{equation} 
where the attempt frequency is given by
\begin{equation} \label{LAMH2}
\Omega=\frac{L}{\xi(T)}\left(\frac{\Delta F}{k_{B}T}\right)^{1/2}\frac{1}{\tau_{\mathrm{GL}}},
\end{equation}
and $\tau_{\mathrm{GL}}=[\pi\hbar/8k_{B}(T^*-T)]$ is the GL relaxation time. Following a development of the energy barrier by Lau $et\ al.$\cite{Lau2001}, we can write $\Delta F(T)$ as, 
\begin{equation} \label{LAMH3}
\Delta F(T)=Ck_{B}T^*\left(1-\frac{T}{T^*} \right)^{3/2},
\end{equation}	
where $C$ is a dimensionless parameter relating the energy barrier for phase slips $F$ to the thermal energy near $T_{onset}$ and is defined as,
\begin{equation} \label{LAMH4}
C\approx0.83\left(\frac{L}{\xi(0)}\right)\left(\frac{R_{q}}{R_{F}}\right).
\end{equation}
Here, $R_{q}=h/4e^{2}=6.45$~k$\Omega$ is the resistance quantum for Cooper pairs and $R_{F}$ the normal state resistance of the entire nanowire~\cite{Cirillo2012}. It can easily be shown that Eqn.~\ref{LAMH4} remains valid for q1D superconducting arrays rather than individual nanowires, subject to the replacement of $R_{F}$ by the $total$ crystal resistance $R_{NS}$ and the crystal length $L$ by the renormalized length $L_{eff}$.

\subsubsection{Quantum phase slips (QPS)}
QPS are expected to become relevant only when $k_{B}T<\Delta(T)$~\cite{Cirillo2012}.  Since {\Tl} is a strong-coupling superconductor, we estimate that QPS should become applicable for $T\lesssim0.9T_p\equiv$~5.3~K: comfortably within the $T_{3D}^{ons}<T<T_p$ range in which we anticipate phase slips. For the QPS contribution, we used the following expression~\cite{Arutyunov2008}:
\begin{equation} \label{QPS1}
R_{QPS}(T)=A_{Q}B_{Q}\frac{R_{q}^{2}}{R_{F}}\frac{L^{2}}{\xi(0)^{2}}\exp\left[-A_{Q}\frac{R_{q}}{R_{F}}\frac{L}{\xi(T)}\right],
\end{equation}
where $A_{Q}$ and $B_{Q}$ are constants.  In a similar manner to the TAPS contribution, we treat our crystals as macroscopic arrays of nanowires and rewrite Eq.~(\ref{QPS1}) in terms of $L_{eff}$ and the normal state resistance $R_{NS}$ of the entire crystal. Finally, the total theoretical $R(T)$ is calculated by considering a parallel combination of the TAPS and QPS components, with an additional quasiparticle contribution $R_{NS}$ as follows,
\begin{equation} \label{Rtotal}
R=(R_{NS}^{-1}+(R_{LAMH}+R_{QPS})^{-1})^{-1}.
\end{equation} 

The solid curves in Fig.~\ref{Fig5} show the least-square fits to our experimental $R_c(T)$ data from 4.9~K to 5.9~K, using Eq.~(\ref{Rtotal}) with the fitting parameters $T_{onset}$, $L_{eff}/\xi(0)$, $A_{Q}$ and $B_{Q}$. $R_{NS}$ = 0.0095 $\Omega$ is used for all the fits. The fitting parameters with the error bars are listed in Table~1.  In q1D superconductors whose resistance is influenced by QPS, $A_{Q}$ is expected to be of order unity, in agreement with our data. Crucially, the pairing temperature remains an unconstrained variable during fitting yet invariably yields values in the range 5.9-6.1~K, consistent with our observed $T_p$.

\begin{table}[htbp]
	\caption{\textbf{Phase slip fit parameters for Tl$_2$Mo$_6$Se$_6$ Sample\#1 (4.9~K -- 5.9~K)}} 
	~Fits are shown in Fig.~\ref{Fig5}.
	\centering 
	\small\addtolength{\tabcolsep}{-1.6pt}
	\begin{tabular}{ccccc} 
		\hline\hline 
		I (mA) & $T^*$~(K) & $L_{eff}/\xi(0)$ & $A_{Q}$ & $B_{Q}$ \\ [0.4ex] 
		\hline 
		0.05 \vline& 5.89$\pm$0.02 \vline& (1.44$\pm$0.12)$\times$10$^{-4}$ \vline& 0.43$\pm$0.04 \vline& (20$\pm$2.2)$\times$10$^{-4}$ \\
		0.10 \vline& 6.02$\pm$0.02 \vline& (1.08$\pm$0.07)$\times$10$^{-4}$ \vline& 0.45$\pm$0.04 \vline& (40$\pm$3.1)$\times$10$^{-4}$ \\ 
		0.20 \vline& 6.16$\pm$0.02 \vline& (7.82$\pm$0.52)$\times$10$^{-5}$ \vline& 0.52$\pm$0.05 \vline& (70$\pm$5.3)$\times$10$^{-4}$ \\ [1ex] 
		0.50 \vline& 6.00$\pm$0.08 \vline& (5.71$\pm$0.36)$\times$10$^{-5}$ \vline& 0.92$\pm$0.04 \vline& (70$\pm$20)$\times$10$^{-4}$ \\ [1ex] 
		\hline 
	\end{tabular}
	\label{table:PdBi2} 
\end{table} 


%

\end{document}